\documentclass{elsart5}

 \usepackage{graphicx}

\usepackage{amssymb}

\begin{document}

\begin{frontmatter}

\title{Interplay between different states in heavy fermion physics}
\author{G. Knebel\corauthref{cor1}}
\corauth[cor1]{}
\ead{georg.knebel@cea.fr}
\author{K. Izawa}
\author{F. Bourdarot}
\author{E. Hassinger}
\author{B. Salce}
\author{D. Aoki}
\author{J. Flouquet}

\address{D\'{e}partement de la Recherche Fondamentale sur la Mati\`{e}re Condens\'{e}e, SPSMS, CEA Grenoble,\\ 17 rue des Martyrs, 38054 Grenoble Cedex 9, France}



\begin{abstract}
Calorimetry experiments under high pressure were used to clarify the interplay between different states such as superconductivity and antiferromagnetism in CeRhIn$_5$, spin density wave and large moment antiferromagnetism in URu$_2$Si$_2$. Evidences are given on the re-entrance of antiferromagnetism under magnetic field in the superconducting phase of CeRhIn$_5$ up to $p_c = 2.5$~GPa where the N\'eel temperature will collapse in the absence of superconductivity. For URu$_2$Si$_2$ measurements up to 10~GPa support strongly the coexistence of spin density wave and large moment antiferromagnetism at high pressures. 
\end{abstract}



\end{frontmatter}


\section{Introduction}
With the recent development in microcalorimetry experiments under extreme conditions [very low temperature ($T$), high pressure ($p$) or strong magnetic field ($H$)] \cite{Salce2000} a new period starts in heavy fermion physics with the possibility of a precise determination of the boundaries between different phases. The first example is given by the recent experiments performed on the heavy fermion compound CeRhIn$_5$ in Los Alamos \cite{Park2006} as well as in Grenoble \cite{Knebel2006}; it leads to state more precisely the interplay between antiferromagnetism (AF) and superconductivity (SC). The spectacular effect is the re-entrance of AF under magnetic field ($H$) inside the SC phase. 

The second example discussed will be URu$_2$Si$_2$; the long standing debate is the duality between the low pressure ($p<p_x\sim 0.5$~GPa) hidden order (HO) phase and the switch to a large moment antiferromagnetism (LMAF) \cite{Mydosh2003,Amitsuka2006,Bourdarot2005}. Recent microcalorimetric measurements up to 12~GPa reveal the coexistence of LMAF and a spin density wave (SDW) above $p_x$. The SDW may be the major component of the HO state below $p_x$ with the exclusion of LMAF since a drop in the carrier number coincides with the entrance in the HO phase.

There are other interesting examples such as the interplay of the SC domain with the different ferromagnetic phases of UGe$_2$ \cite{Pfleider2002} and of URhGe driven by pressure or magnetic field \cite{Levy2005,Huxley2006}, or the collapse of the insulating phase (I) of SmB$_6$ or SmS and the concomitant appearance of long range magnetism \cite{Derr2006,Matsubayashi2006}.

Often, the interplay between AF and another phase (SC, SDW, or I), with a tuning parameter ($\frac{p-p_c}{p_c}$), leads not to a second order phase transition as assumed in the quantum-critical approach (the N\'eel temperature $T_N \to 0$)  but via a first order transition. There are also situations like URu$_2$Si$_2$ or UGe$_2$ where SC can disappear or survive under $p$ on crossing a first order line depending on the $p$ variation of the pairing potential \cite{Flouquet2006}.

In the two chosen examples of CeRhIn$_5$ and URu$_2$Si$_2$, a strong duality exist between the localized and itinerant character of the 4$f$ and 5$f$ electrons. An important part of the puzzle is the Fermi surface (FS) topology (see R. Settai \cite{Settai2006}). A main difference between the two systems is that in CeRhIn$_5$ the number of electron carriers $n_e$ is roughly comparable to the number of the magnetic sites while in URu$_2$Si$_2$ a large decrease of the carrier number (a factor of 3 or 10) is associated to the ordering temperature $T_0$ going from paramagnetic (PM) to  the HO state. 

\section{Competition between AF and SC in CeRhIn$_5$}
\begin{figure}[t]
\begin{center}
\includegraphics[scale =1.3]{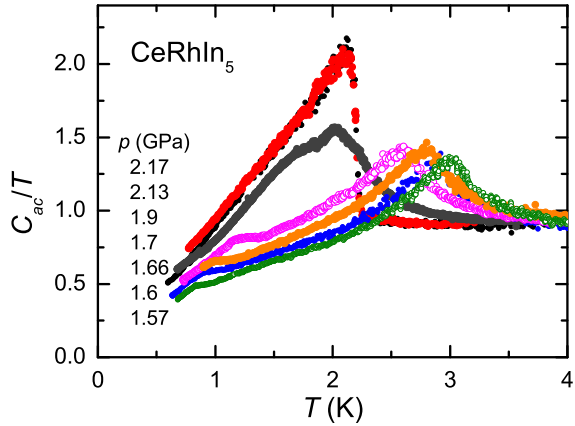}
\caption{At $H=0$, temperature variation of the ratio of the specific heat divided by $T$ for $p$ ranging from 1.57 to 2.17~GPa. Below $p_c^* \sim 1.95$~GPa, the SC anomaly is weak and broad when $T_N > T_c$; AF disappears above $p_c^*$.
}
\label{fig-1}
\end{center}
\end{figure}

The discovery of the 115 cerium family has opened the possibility of a careful study of the interplay between AF and SC. Since the maxima of their N\'eel temperature $T_N$ ($\sim 3.8$~K) and their superconducting temperature $T_c$ ($\sim 2.2$~K) are quite comparable as well as the size of their associated specific heat anomaly \cite{Thompson2001}. Figure \ref{fig-1} illustrates the data of ac microcalorimetry experiments under pressure on CeRhIn$_5$ down to 0.6~K \cite{Knebel2006}. Below 1.5~GPa, only AF anomalies are detected; from 1.5~GPa to $p_c^* \sim 1.95$~GPa so far $T_N(p) > T_c(p)$, on cooling below $T_N$ a broad SC specific heat anomaly is detected at $T_c^C (p)$ with $T_c^C (p)$ rapidly increasing under pressure. However, the detection of the diamagnetic SC shielding by ac susceptibility gives another value of $T_c$, $T_c^\chi (p) > T_c^C (p)$. Clearly the onset of SC appears quite inhomogeneous or  at least quite different from the classical BCS prediction. The recollection of resistivity ($\rho$), susceptibility ($\chi$), and specific heat ($C$) data gives the sequence $T_c^\rho > T_c^\chi > T_c^C$ in the respective $T_c$ determination. Recently it was claimed only on the basis of ac susceptibility measurements that even at $p=0$ CeRhIn$_5$ may present a coexistence of SC and AF with $T_c (p=0) \approx 90$~mK and a superconducting critical field $H_{c2} (0)$ at $T\to 0$ near 500~Oe \cite{Shen2006}. This conclusion deserves careful verifications as lattice imperfections may lead to superconducting behavior not directly related with SC bulk nature unambiguously detected by ac calorimetric experiments. 

By contrast in microscopic NQR measurements, the AF-SC matter below $p_c^*$ appears homogeneous with a single relaxation process and a well characterized AF pattern. However, SC is mainly gapless \cite{Kawasaki2003}. By comparison to ordinary systems showing the coexistence of AF and SC as the borocarbide or Chevrel phases with quite different electronic baths involved in AF (localized spin) and in SC (light itinerant electron), the first novelty of heavy fermion superconductors is that, in zero order, the same electrons are involved in the magnetic and SC properties. The usual statement that SC will not modify AF even for $T_c>T_N$ is clearly precluded \cite{Flouquet2006}. The second novelty is that the magnetic coherence length $\xi_m$ can reach a nanometric scale, far higher than the atomic distance observed in classical AF, due to the proximity of the so-called quantum critical point (QCP); even $\xi_m$ may become comparable to the superconducting coherence length $\xi_0$. 

\begin{figure}[t]
\begin{center}
\includegraphics[scale =0.5]{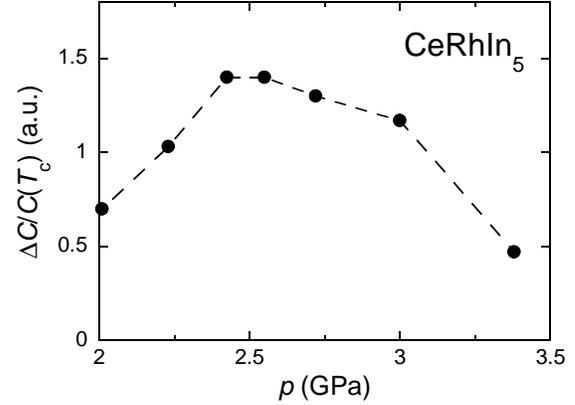}
\caption{Jump $\Delta C/C$ of the Sc specific heat anomaly above $p_c^*$.
}
\label{fig-2}
\end{center}
\end{figure}

Obviously above $p_c^* \sim 1.95$~GPa when $T_c (p)$ becomes higher than $T_N (p)$, AF disappears suddenly at least down to the lowest measurement temperature ($T=0.5$~K). The usual molecular field shape of the SC specific heat anomaly is recovered and there is a perfect coincidence between $T_c^\rho = T_c^\chi = T_c^C$. Furthermore, the size of the SC anomaly $\Delta C/C$ at $T_c$ as a function of $p$ goes through a maximum around $p_c \sim 2.5$~GPa (see fig.~\ref{fig-2}). Extrapolating from CeCoIn$_5$ behavior \cite{Petrovic2001}, this extremum is a signature of the enhancement of the effective mass: the increase of $\Delta C/C$ at $T_c$ near $p_c$ is associated to the fact that the Fermi liquid is far to be established at $p_c$ so the extrapolation of $C/T$ in the normal phase down to $T=0$ will be quite higher than the value of $C/T$ just above $T_c$. Direct evidences of an increase of the effective mass above $p_c^*$ come from dHvA experiments \cite{Shishido2005}. In these high field experiments ($H> 8$~T), it is also claimed that the Fermi surface changes at $p \sim p_c$ supporting the idea that the QCP may be associated with a localization - delocalization transition of the 4$f$ electron. However, the SC properties in low field do not support a discontinuous change of the FS at $p_c$. Thus, we think the strong non-symmetric increase of the effective mass on both sides of $p_c$  reported in the dHvA experiments reflects mainly the transition from AF ordered state to PM state (see below). The link between a change in dHvA frequencies and localization of the 4$f$ electron is not so obvious. In the well documented case of CeRu$_2$Si$_2$ \cite{Flouquet2006}, a change in the itinerancy of the 4$f$ electrons under magnetic field at its pseudo-metamagnetic field $H_m \sim 7.8$~T is unlikely to occur; the high magnetic polarization of the band may lead to a reconstructing of the FS \cite{Daou2006,Miyake2006}. The directly following questions is if the FS at $H=0$ may not vary at $p_c^*$ when AF disappears quasi-discontinuously.
\begin{figure}[t]
\begin{center}
\includegraphics[scale =1.3]{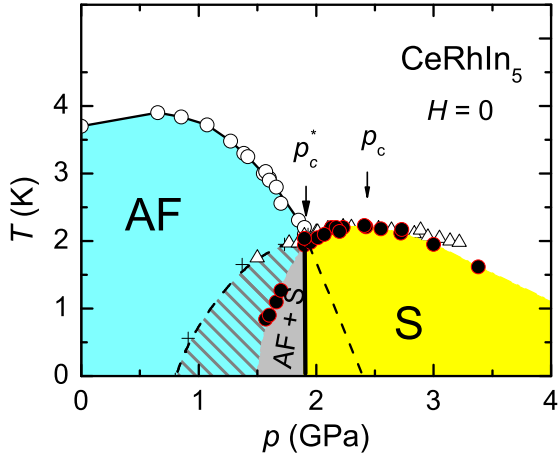}
\caption{($p,T)$ phase diagram of CeRhIn$_5$ at $H=0$, the open ($\circ$) and full ($\bullet$) circles represent the AF and SC anomalies, respectively; ($\triangle$) and ($+$) correspond to $T_c^\chi$ and $T_c^\rho$ (see \cite{Knebel2006}). The dashed line is an extrapolation of $T_N$ in absence of superconductivity.
}
\label{fig-3}
\end{center}
\end{figure}

Above $p_c^*$, also a clear change is found in NQR properties; the nuclear relaxation rate $1/T_1$ follows the well known $T^3$ law below $T_c$ characteristic of unconventional superconductivity with a line of zeros. New careful NQR experiments reported in this conference \cite{Yashima2006}, suggests that the phase diagram is not as shown in Fig.~\ref{fig-3} with a first order transition but there is a tetracritical point at $p_c^*$ with a clear coexistence of a homogeneous regime AF+SC between the purely AF and SC states as discussed in SO(5) theory \cite{Demler2004}. Of course, the next step is to observe  at low temperature directly the end points (quantum critical or first order) $p_{-s}$ and $p_{AF}$ where the SC or AF component of the AF+SC phase collapse, respectively; $p_{-s}$ seems to be near 1.5~GPa, $p_c^* \sim 1.95$~GPa with $T_N(p) =T_c(p)$, and $p_{AF} \sim 2.1$~GPa. 

In a simple picture there is a competing process between the AF and the SC gap. If $T_c(p) > T_N(p)$ (above $p_c^*$), the SC gap is mainly opened on almost the entire $k$ space (outside a restricted domain where a line of zeros may occur); it precludes the further occurrence of AF on cooling. On the other hand below $p_c^*$, a coexisting regime of AF and SC can occur since here only a restricted $k$ space is involved by the occurrence of AF; however, AF can only persist in the pressure range of gapless SC (see discussion in ref.~\cite{Shen2006}).

\begin{figure}[t]
\begin{center}
\includegraphics[scale =1.3]{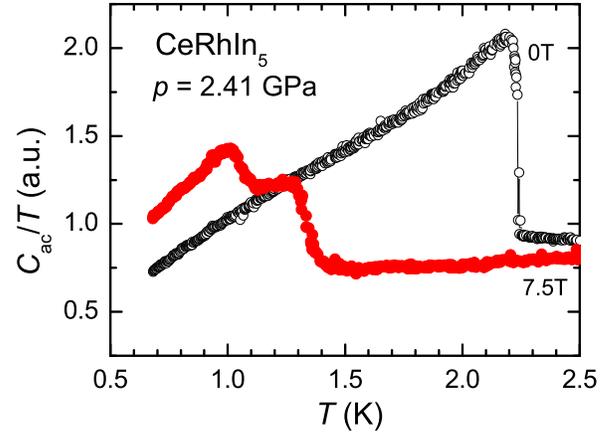}
\caption{$C/T$ versus $T$ for CeRhIn$_5$ at 2.41~GPa slightly below $p_c$ at $H=0$ and $H=7.5$~T$< H_{c2}\sim 10$~T.
}
\label{fig-4}
\end{center}
\end{figure}

\begin{figure}[t]
\begin{center}
\includegraphics[scale =1.3]{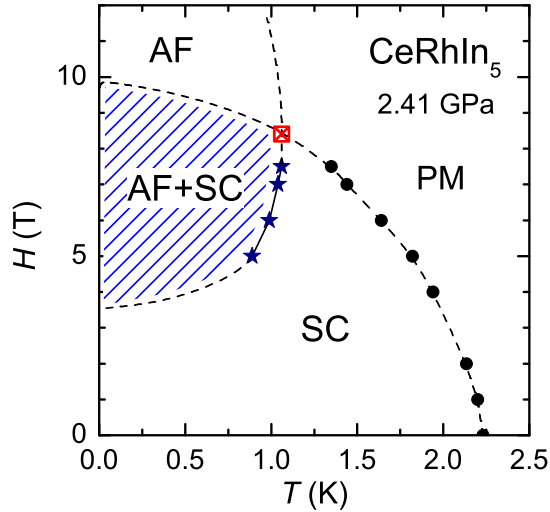}
\caption{Summary of the $H$ re-entrant phenomena (AF+SC) at $p=2.41$~GPa; full points are given by the experiment, the ($\boxtimes$) indicates the point where $T_c(H) =T_N(H)$ (the field where $T_c(H) =T_N(H)$ defines $H_{1,2}$, see text).  Dashed lines are the suspected zone boundaries from other $p$ points. The low field extrapolation of the re-entrant AF+SC phase agrees with the result of ref. \cite{Park2006}.
}
\label{fig-5}
\end{center}
\end{figure}

In this interplay between magnetism and SC, the magnetic field $H$ can reveal new situations by modifying the nature of the interaction itself (switching from antiferromagnetic to ferromagnetic \cite{Flouquet2006}) or also by reversing the relative strength of $T_N$ versus $T_c$. For the AF CeRhIn$_5$, the characteristic field $H_m$ of the AF - PM boundary is very high at ambient pressure, near 50 T along the basal plane \cite{Takeuchi2001}. Without SC the N\'eel temperature is suspected to collapse near $p_c$ and furthermore, from previous studies on heavy fermion compounds \cite{Flouquet2006}, it is believed that $H_m$ may not collapse linearly with pressure at $p_c$ but vanishes only rapidly for $p$ close to $p_c$. As $H_{c2}(0) \sim 10$~T at $p_c^*$ is quite lower than $H_m (p=0)$, $T_c(p)$ for $p_c > p > p_c^*$ will decrease rapidly with $H$ while $T_N(p)$ is almost field independent for $H < H_{c2} (0)$. At a field $H_{1,2}$ when $T_c(p)$ reaches $T_N(p)$, AF is certainly recovered in the SC phase as shown in Fig. \ref{fig-4}. Neglecting the vortices, the coexisting phase of AF+SC under field will correspond to $H_{c2} > H_{1,2}$, i.e. defined by the $H_{c2}$ curve $H_{c2} > H_{1,2}$ and the horizontal line $H = H_{1,2}$ line.  The interesting point is that the coexisting domain occurs far below $H_{1,2}$ (see Fig.~\ref{fig-5}). This clearly proves that the vortex play an important role in the re-entrance of the antiferromagnetism under magnetic field. In the Grenoble experiment, no second transition can be observed for $H<4$~T, whatever is $p>p_c^*$. In Los Alamos data AF is detected from $H=0$ as $p\to p_c^*$ \cite{Park2006}. It was suggested that such a re-entrance of AF is well explained in SO(5) theory \cite{Demler2004}; the magnetism originated from the vortex core can expand on the superconducting coherence length as basically $\xi_m$ is comparable to $\xi_0$. On approaching $p_c$, the coexisting domain shrinks toward $H_{c2} (0)$. This suggests strongly that magnetism may play an important role in the emergence of a new low temperature -- high magnetic field phase in CeCoIn$_5$ \cite{Bianchi2003,Radovan2003} when the field is applied in the basal plane (see NMR contribution of \cite{Kakuyanagi2005,Kumagai2006} and contradictory reports of refs. \cite{Mitrovic2006} and \cite{Young2006}). Very recently the interplay of SC and AF has been studied in CeCo(In$_{1-x}$Cd$_x$)$_5$ \cite{Pham2006} with the confirmation that without SC CeCoIn$_5$ will be near an AF QCP and with also the observation that AF and SC coexist for $x>0.075$ when $T_N$ is larger than $T_c$ but AF collapses so far SC appears at a higher temperature than the expected value of $T_N$ for $x<0.075$.

Now there are converging evidences of the duality between AF and SC as well as the possibility of coexisting phases induced under pressure or magnetic field. The next step is to go further: microscopic informations on the nature of the order parameter in the different regimes (notably on the  incommensurability or commensurability of the AF structure in the AF and SC) and the key role of the vortex matter have to be achieved.

One can speculate if in heavy fermion superconductors a second order QCP exist in the presence of SC \cite{Flouquet2006}. Often, in cerium heavy fermion superconductors like CeIn$_3$, CeRh$_2$Si$_2$, or CePd$_2$Si$_2$ the maximum of $T_N$ is about ten times larger than that of $T_c$; thus the study of the AF and SC boundaries is difficult. The repulsion between AF and SC was one of the main results of CeCu$_2$Si$_2$ (see refs. \cite{Bruls1994,Thalmeier2004}. In  the calorimetric experiment on CePd$_2$Si$_2$ \cite{Demuer2002}, no magnetic anomaly has been detected when $T_c>T_N$ as well as no superconducting anomaly when $T_N > T_c$. However, for CeRhIn$_5$ the maxima of $T_N$ and $T_c$ are of the same order of magnetitude, thus enlightening detailed experiments with fine tuning of pressure are possible. For the other cases very low temperatures are required which give rise to a lack of carefull measurements. In the case CeRhIn$_5$ the very fast drop of $T_N$ at $p_c^*$ gives strong indications for a first order transition from an AF+SC to a pure SC ground state in zero magnetic field. The application of a magnetic field yields to the re-entrance of magnetism and shifts the transition to higher pressures; the re-entrance phase seems to collapse close to the critical pressure $p_c$ where $T_N$ would collapse in absence of SC. However, the appearance of SC hides  this magnetically critical regime.

\section{URu$_2$Si$_2$: Competition of Hidden Order (mainly SDW) and LMAF}

\begin{figure}[t]
\begin{center}
\includegraphics[scale =1.3]{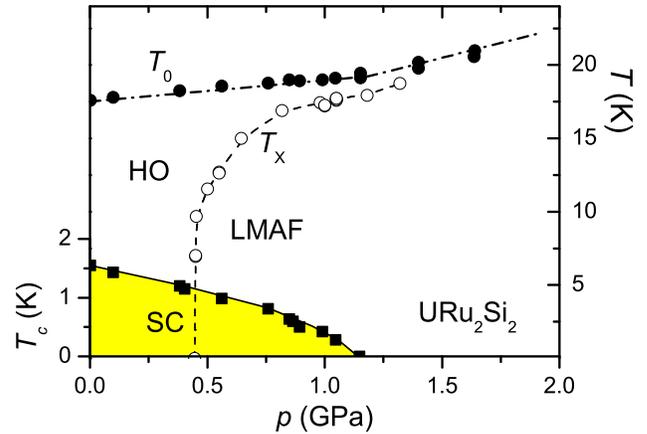}
\caption{($p,T$) phase diagram of URu$_2$Si$_2$ from ref.\cite{Schmidt1993} and \cite{Bourdarot2005}. Open circles ($\circ$) gives the first order ($p_x,T_x$) line as detected by neutron scattering experiments. The uncertainties in the determination of the transition temperatures between different experiments can go up to 1K for $T \sim 20$~K.
}
\label{fig-6}
\end{center}
\end{figure}

The discovery in Nagoya \cite{Montoyama2003} that a first order transition at $p_x=0.5$~GPa occurs in URu$_2$Si$_2$ at low temperature between the hidden order (HO) phase and large moment antiferromagnetism (LMAF) leads us to revisit the phase diagram of URu$_2$Si$_2$ as described in Fig.~\ref{fig-6}. From previous works \cite{Schmidt1993}, the proposal is that the hidden order phase with an ordering temperature $T_0$ is switched to LMAF which may order at $T_N$ higher than $T_0$ at high pressure. The crossing of the $T_0(p)$ and $T_N(p)$ lines seems to occur for $p^* \sim 1.2$~GPa. Thus $p^*$ may be the end point of the $(T_x,p_x)$ dashed line. Another interesting observation is the disappearance of SC near $p^*$; LMAF and SC are antagonist ($T_c \to 0$ when the LMAF volume fraction $f_{LMAF} \to 1$). 

Evidences of the new high pressure phase (LMAF) were first given in ref. \cite{Amitsuka1999} with, however, the preliminary conclusion that the switch from HO to LMAF corresponds to $p\sim p^*$. Further neutron scattering experiments performed under better hydrostatic pressure conditions clarified that the HO to LMAF transition at $T\to 0$~K coincides with $p_x$ \cite{Bourdarot2005}. The nature of the HO phase is still under debate, notably on the intrinsic origin of the detected tiny sublattice magnetization $M_0 \sim 0.03 \mu_B$ at ambient pressure. NMR as well as $\mu$SR experiments \cite{Matsuda2001,Amato2004} favor the extrinsic origin \cite{Mydosh2003,Amitsuka2006} with the underlying idea that pressure gradients near lattice imperfections may stabilize a tiny percentage of LMAF in the HO state. The complete disappearance of the LMAF phase (fraction $f_{LMAF}$) may occur only for a slightly negative pressure ($p_{x}- 0.7$~GPa) while the HO phase may only disappear above ($p_{x}+ 0.7$~GPa). However, a neutron scattering analysis supports an intrinsic origin of the magnetic moment in the HO state \cite{Bourdarot2005}; within this frame a phenomenological model was developed on the basis that the primary HO is a spin density wave which can drive an extra tiny moment on the U site \cite{Mineev2005}.

\begin{figure}[t]
\begin{center}
\includegraphics[scale =.5]{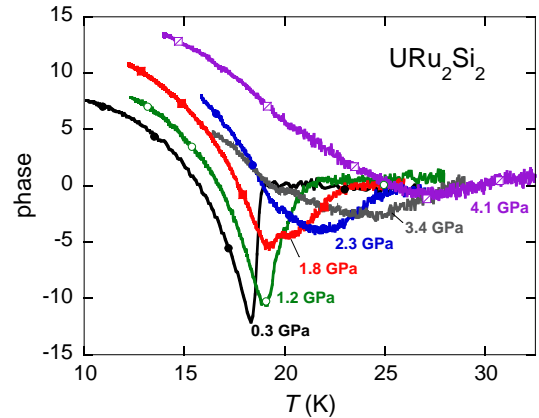}
\caption{Temperature variation of the phase of the ac calorimetric signal at different pressures.}
\label{fig-7}
\end{center}
\end{figure}\begin{figure}[t]
\begin{center}
\includegraphics[scale =.5]{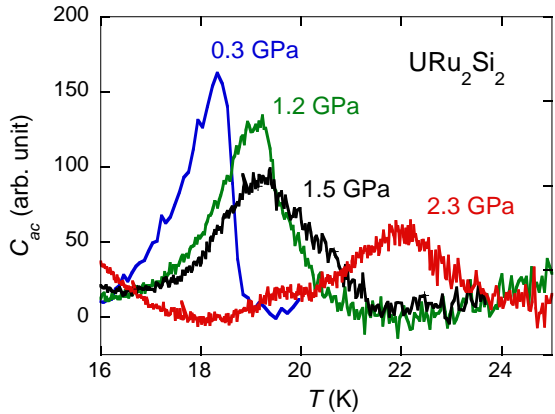}
\caption{Temperature dependence of the inverse module of the measured ac thermoelectric signal at different pressures.
}
\label{fig-8}
\end{center}
\end{figure}

\begin{figure}[t]
\begin{center}
\includegraphics[scale =1.3]{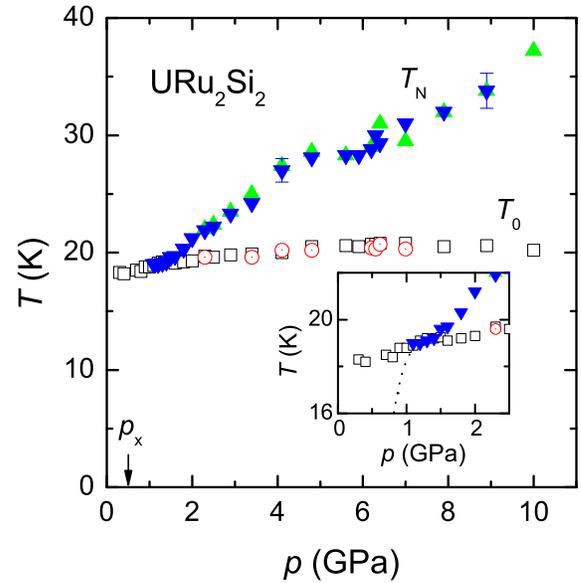}
\caption{Characteristic temperatures $(T_0,T_N)$ of URu$_2$Si$_2$ as detected by microcalorimetry under high pressure. Above 1.5~GPa, the lower temperature seems to be the continuation of the $T_0$ line detected below 1.1~GPa; the higher one is suspected to be $T_N$ for LMAF. 
}
\label{fig-9}
\end{center}
\end{figure}

The relevance of nesting at $T_0$ was already clear two decades ago in URu$_2$Si$_2$ \cite{Maple1986,Schoenes1987}; confirmations were recently given by thermal transport measurements \cite{Behnia2005,Sharma2005}. Furthermore it was stressed that this SDW may be described by a BCS approach. Extrapolating to the previous CeRhIn$_5$ description, when $T_0 > T_N$ LMAF is excluded. To test the coexistence of SDW and LMAF under pressure, new ac calorimetry experiments were performed up to 12~GPa. Figures \ref{fig-7} and \ref{fig-8} represent the phase and inverse of the module of the measured thermoelectric signal, which is, in first order, proportional to the specific heat \cite{Derr2006}. Fig.~\ref{fig-9} displays the high pressure phase diagram deduced from these measurements. The main result is that SDW survives inside the LMAF phase as found for CeRhIn$_5$ for AF and SC in a narrow pressure range. Up to 1.1~GPa, both ac calorimetric responses in phase and module reproduces the $p=0$ behavior. For $p> 1.1$~GPa, the signal slightly above $T_0$ becomes  broadened as the second contribution slowly grows with $p$. Above 1.5~GPa two separated anomalies emerge. The high temperature one is assumed to be associated with the appearance of LMAF at $T_N$; the low temperature one seems to be the continuation of $T_0$.

Of course, at low temperature no specific heat anomaly can be detected along the $(T_x, p_x)$ line as the discontinuous changes are in the volume and sublattice magnetization. As pictured on fig.~\ref{fig-9}, it was expected to observe a calorimetric signal on warming above 10~K, where the $(T_x, p_x)$ line slowly changes its slope but we were unable to detect any specific heat contribution. Clearly in an inhomogeneous model, so far $f_{LMAF} \to 1$ ($p \sim 1.5$~GPa), $T_N$ cannot be discriminated from $T_0$. 

The persistence of nesting in all the pressure range is in excellent agreement with the stability of the Fermi surface \cite{Nakashima2003} and also the shape of the resistivity anomaly close to $T_0$ above $p_x$ \cite{Schmidt1993,McElfresh1987}. One scheme is that, at low pressure, the molecular field acting on the localized spin (which is the combined result of an exchange among localized spin and of an interaction via the polarization of the quasi-particles) is too weak for the appearance of LMAF while nesting is quite favorable \cite{Miyake}. At the critical pressure $p_x$, its strength become critical and LMAF is the new stable solution as predicted for the induced magnetism on singlet crystal field ground state. 

Another open question is if $p_x$ does not mark a tiny discontinuous change of the valence inducing differences in spin and orbital 5$f$ components. The URu$_2$Si$_2$ case may be similar to the YbInCu$_4$ \cite{Mito2003,Kogama2005} or SmB$_6$ \cite{Derr2006} excitations with a first order point between an intermediate valent state and long range magnetism and also a residual phase separation around $p_x$.

There is no doubt that SC is linked to HO. When  $f_{LMAF} \to 1$, SC disappears. All experiments agree with the disappearance of SC around $p^*$. Recent ac susceptibility measurements suggest that $T_c$ collapse discontinuously at $p_x$ \cite{Sato2006}, while a dc magnetization probe \cite{Tenya2005} indicates that $T_c$ may reach zero only near $p^*$ as reported already by pressure and field resistivity studies. Ac calorimetry experiments are required to clarify this issue. In resistivity experiments it was also found that the $A$ coefficient of the $T^2$ term drops by a factor of 4 between ambient pressure and 2 GPa \cite{Schmidt1993,McElfresh1987} in agreement with the image of the development of a large molecular field below $T_N$. It is worthwhile to mention that in resistivity measurements even above $p_x$ the derived $H_{c2}$ curves follow the pressure variation of the effective mass $m^\star$ assuming the usual proportionality $A \propto {m^\star}^2$ \cite{Schmidt1993}. As has been observed for the sublattice magnetization in detail, in an inhomogeneous description with a fraction of HO in LMAF and reciprocally, the extrinsic properties are coupled to the bulk properties.  

An interesting problem is the behavior of URu$_2$Si$_2$ under magnetic field. From earlier Hall effect measurements \cite{Bakker1993} and their recent extension to Rh doping \cite{Oh2006} it was clear that the closing of the SDW gap at $H_\Delta \sim 38$~T leads to recover a high carrier number as generally found in metallic heavy-fermion antiferromagnets. The high magnetic field phase diagram of URu$_2$Si$_2$ reflects the interplay between nesting and localized magnetism. The cascade of high field induced ordered phase (II, III, V) in ref. \cite{Kim2003}
is directly linked to the drastic increase in the carrier number; the polarized paramagnetic phase IV is quite similar to that found in CeRu$_2$Si$_2$. Of course, different fields can modulate the interplay but the disappearance of nesting governs the entrance in the PM state: $H_\Delta \sim H_M$. 

For the HO phase, no clear evidence has been found on another ordered state than SDW. However, even the nesting vector of the SDW is not determined; it is only suspected to be the (1,0,0) wavevector of the LMAF. A yet unclear problem is the link between the main spin excitations $\Delta_{1,0,0}$ and $\Delta_{1.4,0,0}$ at the wavevectors (1,0,0) and (1.4,0,0) \cite{Bourdarot2005,Broholm1991,Wiebe2004}: are they coupled via a dispersion relation or the result of two different origins? At zero pressure in magnetic field, $\Delta_{1,0,0}$ increases and seems to reach $\Delta_{1.4,0,0}$ at $H_M$ \cite{Bourdarot2003a}. This supports the idea that above $H_M$ the 5$f$ electrons are governed by local fluctuations. In zero magnetic field $\Delta_{1,0,0}$ appears to collapse for $p>p_x$ while $\Delta_{1.4,0,0}$ increases \cite{Bourdarot2003b}.

With this new exploration of the ($p,T$) phase diagram of URu$_2$Si$_2$, an evidence is given of the coexistence of LMAF with SDW at high pressure. Soon, x-ray
scattering experiments under $p$ will be realized to confirm if LMAF is set up at $T_N$ above $T_0$ for $p>1.5$~GPa. Extended combined studies under extreme conditions (with large $T,H,p$ scans) will also clarify the magnetic and electronic duality of the localized and itinerant component of the 5$f$ electrons.

\end{document}